\documentclass[preprint,numafflabel]{elsarticle}
\usepackage[top=2cm, bottom=2cm, outer=2.5cm, inner=2.5cm]{geometry}
\usepackage[english]{babel}
\usepackage{lineno,hyperref}
\usepackage{multicol}

\usepackage{mathtools}
\usepackage{amsmath}
\usepackage{amssymb}
\usepackage{bbm} 
\usepackage{widetext} 
\usepackage{amsthm}
\usepackage{multirow}

\usepackage{graphicx}
\usepackage{caption}
\usepackage{array}
\usepackage{float}
\usepackage{ragged2e}
\usepackage{graphbox} 
\usepackage{booktabs} 
\usepackage{colortbl} 
\usepackage{soul}
\usepackage[usernames,dvipsnames,svgnames,table]{xcolor}
\usepackage{hyperref}
\newcommand{\mini}{\scriptscriptstyle}
\DeclareMathOperator{\sign}{sign}
\DeclareMathOperator{\Imag}{Im}

\journal{...}
\modulolinenumbers[5]

\begin{document}


\begin{frontmatter}



\title{Physical significance of generalized boundary conditions: \\an Unruh-DeWitt detector viewpoint on $\text{AdS}_2 \times \mathbb{S}^2$}
\author{Lissa de Souza Campos\footnote{lissa.desouzacampos@unipv.it}$^{\dagger\star}$}
\author{Claudio Dappiaggi\footnote{claudio.dappiaggi@unipv.it}$^{\dagger\star}$}
\author{Luca Sinibaldi\footnote{luca.sinibaldi01@universitadipavia.it}$^{\dagger\star}$}
\address{$^{\dagger}$Dipartimento di Fisica, Universit\`a degli Studi di Pavia, Via Bassi, 6, 27100 Pavia, Italy}
\address{$^{\star}$Istituto Nazionale di Fisica Nucleare -- Sezione di Pavia, Via Bassi 6, 27100 Pavia, Italy}

\begin{abstract}
	On $\text{AdS}_2 \times \mathbb{S}^2$, we construct the two-point correlation functions for the ground and thermal states of a real Klein-Gordon field admitting generalized $(\gamma,v)$-boundary conditions. We follow the prescription recently outlined in \cite{Campos2022byi} for two different choices of secondary solutions. For each of them, we obtain a family of admissible boundary conditions parametrized by $\gamma\in\left[0,\frac{\pi}{2}\right]$. We study how they affect the response of a static Unruh-DeWitt detector. The latter not only perceives variations of $\gamma$, but also distinguishes between the two families of secondary solutions in a qualitatively different, and rather bizarre, fashion. Our results highlight once more the existence of a freedom in choosing boundary conditions at a timelike boundary which is greater than expected and with a notable associated physical significance. 
\end{abstract}


\begin{keyword}
  Generalized Robin boundary conditions \sep Unruh-DeWitt detector \sep Klein-Gordon field \sep $\text{AdS}_2 \times \mathbb{S}^2$
\end{keyword}

\end{frontmatter}

\begin{multicols}{2}


\section{Introduction}


Since the introduction of Unruh-DeWitt particle detectors in quantum field theory \cite{Unruh1976db,dewitt1979quantum}, they have been employed to probe a plethora of features of quantum fields and of spacetimes, stretching from picking out properties of quantum states \cite{Ng2021enc} to identifying topological aspects of underlying backgrounds \cite{Ng2017iqh}. On a globally-hyperbolic spacetime with a timelike boundary, they are particularly sensible to the choice of an underlying boundary condition for the matter fields: phenomena such as the divergence of physical observables \cite{deSouzaCampos2021awm,Namasivayam2022bky} and the anti-Hawking effect \cite{Henderson2019uqo, DeSouzaCampos2020ddx} depend crucially on the boundary condition set. This is consonant with the fact that different boundary conditions are associated to different quantum states, to different dynamics, to different {\em physics} \cite{Ishibashi2003jd}.

In this arena we have recently shown that within free, scalar quantum field theories on static, curved spacetimes with a timelike boundary one may take into account generalized $(\gamma, v)$-Robin boundary conditions in the construction of physically meaningful states \cite{Campos2022byi}. These arise whenever the underlying dynamics can be reduced thanks to the background isometries to a second order differential equation on the half line and they are characterized by two data: a fixed parameter $\gamma\in\mathbb{R}$ and a so-called secondary solution $v$. For each admissible pair $(\gamma, v)$, one can construct a two-point function both for the ground state and also for arbitrary thermal states. All these correlation functions are physically sensible since, per construction, they enjoy the local {\em Hadamard} property, see \cite{Khavkine:2014mta}. In \cite{Campos2022byi}, we have discussed the example of a wave propagating on the two-dimensional half-Minkowski spacetime showing that an Unruh-DeWitt detector can discern not only the value of $\gamma$, but also the choice of secondary solution $v$. Here, we elaborate on this idea and we provide, in full detail, a full-fledged example that demonstrates how the hidden freedom in the mode expansion of the scalar field, which is associated to the choice of $v$, influences the rate of transition of an Unruh-DeWitt particle detector.

In this work, we consider a real, free, scalar field $\Psi$ with mass $m_0$ on $M\equiv\text{AdS}_2 \times \mathbb{S}^2$ spacetime. This background approximates the near horizon geometry of an extremal black hole with unit charge and it is known as the Bertotti-Robinson solution of Einstein-Maxwell equations, see e.g. \cite{Conroy2021aow} but also \cite{Bertotti, Robinson}. For $t\in\mathbb{R}$ and $r\in(0,\infty)$, its line-element can be written as
\begin{equation}
\label{eq: line-element}
  ds^2 = \frac{1}{r^{2}}\left(-dt^2 + dr^2 + r^2 d\mathbb{S}^{2}(\theta,\varphi)\right),
\end{equation}
where $d\mathbb{S}^{2}(\theta,\varphi)$ is the standard line-element on the unit $2$-sphere. Observe that the manifold possesses a timelike conformal boundary at $r=0$. On top of $M$ we consider a real scalar field $\Psi:\text{AdS}_2\times \mathbb{S}^2\rightarrow\mathbb{R}$ whose dynamics is ruled by the Klein-Gordon equation
 \begin{equation}
   \label{eq: KG}
 P\Psi : = (\Box - m_0^2)\Psi=0,
 \end{equation}
where $\Box$ is the D'Alembert wave operator built out of Equation \eqref{eq: line-element}, while $m_0$ is the mass parameter.

In Section \ref{sec: The Klein-Gordon field}, we recollect the main ingredients necessary to construct the two-point functions for ground and thermal states for $\Psi$. The procedure follows the prescription outlined in \cite{Campos2022byi} and it generalizes the results obtained in \cite{Dappiaggi2016fwc} by considering two different families of generalized $(\gamma, v)$-boundary conditions at the conformal boundary, dubbed $(\gamma, v_\kappa)$ with $\kappa=1,2$. The analysis for $\kappa=2$ coincides with that of \cite{Dappiaggi2016fwc}, while for $\kappa=1$ it yields novel two-point functions. Nonetheless, since also in this case the overall analysis is structurally identical to that for $\kappa=2$, we do not dwell into many details and we limit ourselves to sketching the main steps of the construction.

Subsequently, in Section \ref{sec: The transition rate} we obtain an explicit expression for the transition rate of a static Unruh-DeWitt detector interacting either with the ground or with a thermal state of a Klein-Gordon field. We scrutinize the response of the detector to grasp if and how it is affected by the choice of boundary condition. In particular, we focus on the consequences of the freedom in selecting a secondary solution in the construction of Section \ref{sec: The Klein-Gordon field}. In Section \ref{sec: Summary}, we summarize the main results of this work.


\section{Dynamics and Boundary Conditions}
\label{sec: The Klein-Gordon field}


Working with the coordinates of Equation \eqref{eq: line-element}, a real Klein-Gordon field $\Psi$ that solves Equation \eqref{eq: KG} can be decomposed as
\begin{equation*}
	\Psi(t,r,\theta,\varphi) = e^{-i\omega t} \psi(r) Y_\ell^m (\theta,\varphi),
\end{equation*}
where $Y_\ell (\theta,\varphi)$ are the spherical harmonics on the $2$-dimensional unit sphere. Setting $\lambda=-\ell(\ell+1)$, the function $\psi(r)$ solves the radial equation
\begin{align}
\label{eq: the radial eq}
\mathbf{L} \psi(r) :=
&\frac{d^2 \psi(r)}{dr^2} + \left[ \omega^2 + \frac{\lambda-m_0^2}{ r^2} \right]\psi(r)= 0.
\end{align}


\subsection{The radial equation}


\noindent In the following we assume that $m_0\geq 0$ and we introduce the auxiliary quantities
\begin{align}
\label{eq: nu and p}
  &\nu := \frac{1}{2}\sqrt{1-4\lambda  + 4m_0^2 } \quad \text{and}\quad
  p:= \sqrt{\omega^2},
\end{align}
subject to the constraint $\nu > 0, \,\forall \ell\geq0$. Equation \eqref{eq: the radial eq} is a Sturm-Liouville problem with eigenvalue $\omega^2$, which admits a basis of solutions written in terms of Bessel functions of first and second kind:
\begin{subequations}
\label{eq: basis y1 and y2}
\begin{align}
  y_1(r) &=\sqrt{r}  J_\nu (p r), \\
  y_2(r) &=\sqrt{r}  Y_\nu (p r).
\end{align}
\end{subequations}

Following \cite{Zettl:2005}, it is convenient to work on the Hilbert space $L^2((0,\infty), dr)$ and allow $p$ to take a priori any complex value. Hence, according to Weyl's endpoint classification $r\to\infty$ is a limit point, see also \cite{Dappiaggi2016fwc} for a succinct summary of the nomenclature. As a matter of fact, assuming that $\Imag p\neq 0$, the most general solution of Equation \eqref{eq: the radial eq} lying in $L^2((c,\infty),dr)$, $\forall\, c\in (0,\infty)$ can be written in terms of Hankel functions of first and second kind:
\begin{align*}
  \psi_{\mini{\infty}}(r)  =\sqrt{r}  \left[ H^{\mini{(1)}}_\nu(p r) \Theta(\Imag p)  +  H^{\mini{(2)}}_\nu(p r) \Theta(-\Imag p) \right].
\end{align*}
On the other hand, it turns out that $r=0$ is a limit circle point if $\nu\in(0,1)$, since both $y_1,y_2\in L^2((0,c^\prime),dr)$, $\forall\, c^\prime\in(0,\infty)$, as one can infer from the following asymptotic expansion close to the origin:
\begin{align*}
  |y_1(r)|^2 &\overset{r\rightarrow 0}{\sim}r^{1+2\nu} , \\
  |y_2(r)|^2 &\overset{r\rightarrow 0}{\sim}r^{1-2\nu} .
\end{align*}
Observe that
\begin{equation}\label{Eq: mass range}
	\nu  \in[0,1) \iff \ell = 0 \text{ and } m_0^2\in\left[-\frac{1}{4}, \frac{3}{4}\right).
	\end{equation}
Still following the theory of Sturm-Liouville, this entails that only the $\ell=0$ mode calls for a boundary condition at $r=0$, provided that the mass lies in the range individuated in Equation \eqref{Eq: mass range}. When $\nu\geq1$, no boundary condition needs to be imposed at both ends and since this case is not of interest in this work, we shall not consider it further.


\subsection{Generalized $(\gamma, v)$-boundary conditions}


At the limit circle endpoint, $r=0$, we impose generalized $(\gamma,v)$-boundary conditions as introduced in \cite{Campos2022byi}. These identify self-adjoint extensions of the radial operator $\mathbf{L}$ in $L^2((0,\infty),dr)$ and they are fully characterized in terms of a parameter $\gamma\in\mathbb{R}$, of the principal solution, $u$, and of a secondary solution, $v$, which we introduce in the following. The principal solution at $r=0$ reads
\begin{align*}
&u := \sqrt{r} J_\nu (p r)  = y_1(r).
\end{align*}
This is unambiguously identified by the condition that, for any solution $v$ of Equation \eqref{eq: the radial eq} such that $v\neq \lambda u$, $\lambda\in\mathbb{C}$, then $\lim\limits_{r\to 0}\frac{u}{v}=0$.  On the contrary choosing a secondary solution at $r=0$, linearly independent from $u$, is fully arbitrary. In the following we consider two possible choices $v_\kappa$, $\kappa=1,2$:
\begin{subequations}
	\label{eq: sec sol v1 v2}
\begin{align}
v_1 :=& \,p \sqrt{r} Y_\nu (p r) \\  v_2 :=& - p^{2\nu}  \sqrt{r}  J_{-\nu} (p r).
\end{align}
\end{subequations}
In terms of the basis given in Equation \eqref{eq: basis y1 and y2}, they can be written as
\begin{align*}
v_1 :=& \, p \, y_2(r), \\  v_2 :=& -p^{2\nu}  \left[ \cos(\pi\nu ) y_1(r) - \sin(\pi\nu )y_2(r) \right].
\end{align*}
For $ \kappa\in\{1,2\}$ and $\gamma \in \mathbb{R}$, it follows that the solution
\begin{align*}
  \psi_{\kappa} := \cos(\gamma)u -\sin(\gamma) v_{\kappa} ,
\end{align*}
satisfies
\begin{align}
\label{eq: robin bc}
  &\cos(\gamma)W_r[\psi_{\kappa},u] - \sin(\gamma)W_r[\psi_{\kappa},v_{\kappa}]  = 0 ,
\end{align}
where $W_r$ denotes the Wronskian $W_r[\psi_{\kappa},u]:=\psi_{\kappa}\partial_r u- u \partial_r\psi_\kappa$.
If Equation \eqref{eq: robin bc} holds true, we say that $\psi_\kappa$ satisfies a $(\gamma,v_\kappa)$-boundary condition at $r=0$.

\subsection{The radial Green function}


In the next step in our analysis we construct the Green function of the radial equation. To this end, it is convenient to rewrite the solution at infinity as a linear combination of $u$ and $v_\kappa$:
\begin{align*}
\psi_{\mini{\infty}} &= a_\kappa u + b_\kappa v_{\kappa}.
\end{align*}
A direct computation yields
\begin{align*}
& a_1 = 1,\\
& a_2 = 1 +  i\sign(\Imag p) \cot(\pi\nu ),\\
& b_\kappa= \frac{i\sign(\Imag p)}{\Lambda_\kappa},\quad\kappa=1,2,
\end{align*}
where $\Lambda_1 := p$, and $\Lambda_2 := p^{2\nu}\sin(\pi\nu )$. Since $W_r[J_\nu(pr),Y_\nu(pr)]=\frac{2}{\pi r}$, it descends
\begin{align*}
W_r[\psi_{\kappa}, \psi_{\mini{\infty}}] &= ( b_{\kappa}\cos(\gamma) + a_\kappa\sin(\gamma) )   \frac{2\Lambda_\kappa}{\pi }.
\end{align*}
Therefore, following \cite{greenBook}, if $\mathbf{L}$ is the operator as per Equation \eqref{eq: the radial eq}, the radial Green function, implicitly defined by
\begin{align*}
(\mathbf{L}\otimes \mathbbm{1}) G_\kappa(r,r') = ( \mathbbm{1}\otimes\mathbf{L}) G_\kappa(r,r') =\delta(r-r'),
\end{align*}
reads
\begin{align}
\label{eq: radial Green bessel kappa}
G_\kappa(r,r')
         &= \frac{\pi}{2 \Lambda_\kappa} \frac{ \psi_{\kappa}(r_<) \psi_{\mini{\infty}}(r_>)}{ ( b_{\kappa}\cos(\gamma) + a_\kappa\sin(\gamma) )},
\end{align}
where $\psi_{\kappa}(r_<) \psi_{\mini{\infty}}(r_>)$ is a shortcut notation for the integral kernel $\Theta(r-r^\prime)\psi_{\kappa}(r)\psi_{\mini{\infty}}(r^\prime)+\Theta(r^\prime-r)\psi_{\mini{\infty}}(r)\psi_{\kappa}(r^\prime)$.

\subsection{Symmetries of the radial Green function}


Seeing $G_\kappa$, defined in Equation \eqref{eq: radial Green bessel kappa}, as a function either of $\lambda = p^2$ or $\sqrt{\lambda} = p$, we find that
\begin{align}
&G_1(\overline{p}) = \overline{ G_1(p)}, \text{ but } G_1(\overline{\lambda} ) \neq \overline{ G_1(\lambda)}; \label{eq: radial Green bessel kappa complex conjugation G1} \\
&G_2(\overline{\lambda} ) = \overline{ G_2(\lambda)}, \text{ but } G_2(\overline{p}) \neq \overline{ G_2(p)}. \label{eq: radial Green bessel kappa complex conjugation G2}
\end{align}
\noindent We remark that in the subsequent analysis, particularly in Equations \eqref{eq: radial Green bessel kappa complex conjugation G1} and \eqref{eq: radial Green bessel kappa complex conjugation G2}, the map $\lambda \mapsto \overline{\lambda}$ corresponds to, respectively, $p\mapsto \overline{p}$ and $p\mapsto -\overline{p}$. In this last case, one chooses once and for all $\sign(\Imag p )\in\{-1,+1\}$.


\subsection{Bound States}


When choosing a boundary condition in Equation \eqref{eq: robin bc}, we consider as {\em admissible} the values of $\gamma$ for which there does not exist $\omega_b\in\mathbb{C}$ such that $\lim\limits_{\omega\rightarrow\omega_b}G_\kappa$ diverges. In other words, at a physical level, we rule out those boundary conditions for which $G_\kappa$ has {\em bound state} frequencies $\omega_b$. The case for $\kappa=1$ is analogous to the computation on Minkowski spacetime in spherical coordinates as in \cite[Pg.66]{DeSouzaCampos2022wsp}, while the case for $\kappa=2$ has been studied in detail in \cite[Pg.10]{Dappiaggi2016fwc}. Here, we report the final result: neither $G_1$ nor $G_2$ possesses bound state frequencies for $\gamma\in\left[0,\frac{\pi}{2}\right]$. Henceforth we restrict our attention to this interval. Observe that $\gamma=0$ corresponds to the usual Dirichlet boundary condition. It is tempting to claim that, if $\gamma=\frac{\pi}{2}$, we are instead considering a Neumann boundary condition. This is indeed an admissible nomenclature, but the reader should keep in mind that the solution at $r=0$ of the radial equation depends only on the choice of the secondary solution, contrary to the case $\gamma=0$ in which only $u$ plays a r\^{o}le. Hence there is no universal way to select a Neumann boundary condition, inasmuch as there does not exist a unique criterion to single out a secondary solution for Equation \eqref{eq: the radial eq}.


\subsection{The resolution of the identity}


In order to construct the two-point correlation function of the Klein-Gordon field $\Psi$ with arbitrary $(\gamma,v)$-boundary condition, a last necessary ingredient is the so-called resolution of the identity for the radial equation. Following \cite[Ch.7]{greenBook}, this reads
\begin{align}
\delta(r,r') & = -\frac{1}{2\pi i}\oint\limits_{C_\kappa^\infty} G_\kappa(x,x')dp^2  \nonumber \\ & = \int\limits_{-\infty}^{\infty} p\, dp \, \widetilde{\psi}_\kappa(x,x^\prime)\label{eq: resolution identity bessel example 2}
\end{align}
where the contours differ if $\kappa=1,2$ and they are illustrated in \cite[Fig.1]{Campos2022byi}, while the integrand $\widetilde{\psi}_\kappa(x,x^\prime)$ has been computed in \cite{DeSouzaCampos2022wsp}. Since its explicit form is not of relevance in this work, we shall not report it, although, for later convenience, we highlight that, whenever we restrict the attention to real, positive frequencies $p=\omega$,

%

\begin{align*}
\widetilde{\psi}_\kappa(r,r') &=\frac{ \psi_{\kappa}(r_<)\psi_{\kappa}(r_>) }{\mathcal{N}_\kappa},
\end{align*}
where
\begin{align*}
\mathcal{N}_1 &: =2(\cos^2(\gamma)+ p^2 \sin^2(\gamma)),\\
\mathcal{N}_2 &: =2(\sin^2(\gamma) p^{4 \nu } + \sin(2\gamma) \cos (\pi  \nu ) p^{2 \nu }+\cos^2(\gamma)).
\end{align*}
%
%


\subsection{Two-point correlation functions}


In the preceding analysis we have individuated all the ingredients necessary to construct the two-point correlation function both of a ground and of a thermal state for a real Klein-Gordon field whose mass is such that one can endow it with physically admissible boundary conditions, {\it i.e.}, $m_0^2\in\left[-\frac{1}{4}, \frac{3}{4}\right)$. As in the previous sections, we shall not dwell into a detailed construction which has been already outlined in \cite{DeSouzaCampos2022wsp}, but we limit ourselves to recalling the final result

\begin{widetext}
\begin{align}
\label{eq: two point function ground and thermal psi}
\omega_{2,\kappa}(t,r,\theta,\varphi,t',r',\theta',\varphi') =\lim_{\varepsilon\to 0^+}\sum\limits_{\ell=0}^\infty \sum\limits_{m=-\ell}^\ell \int_0^\infty d\omega\, T_{\beta,\varepsilon}(t-t')\widetilde{\psi}_\kappa(r,r') Y_\ell^m  (\theta,\varphi)\overline{Y_\ell^m  (\theta',\varphi')},
\end{align}
\end{widetext}
\noindent where
\begin{equation}
\label{eq: two point function ground and thermal time function}
T_{\beta,\varepsilon}(t-t') := \left[ \frac{e^{-i \omega (t-t' - i \varepsilon)}}{1-e^{-\beta \omega}} + \frac{e^{i \omega (t-t' + i \varepsilon)}}{e^{\beta\omega}-1}\right].
\end{equation}
For finite $\beta$, Equation \eqref{eq: two point function ground and thermal psi} together with Equation \eqref{eq: two point function ground and thermal time function} characterize a thermal state at temperature $1/\beta$. In the limit $\beta\rightarrow\infty$, we obtain instead the two-point correlation function of the ground state:
\begin{equation}
T_{\infty,\varepsilon}(t-t') := e^{-i \omega (t-t' - i \varepsilon)}.
\end{equation}
In the next section we shall use this class of two-point correlation functions to show that the choice of different secondary solutions associated to Equation \eqref{eq: the radial eq} is not a mere mathematical exercise, but it bears strong physical consequences.


\section{The transition rate}
\label{sec: The transition rate}


In this section we consider an Unruh-deWitt detector, namely a spatially localized two-level system that interacts with the underlying scalar field via a monopole type interaction. Here we are interested in computing and analyzing the transition rate when the detector follows static trajectories on $\text{AdS}_2 \times \mathbb{S}^2$. The detector can be initially either in the ground-state $|0\rangle$ or in the excited state $|\Omega\rangle$, see \cite{Unruh1976db, dewitt1979quantum}. This information is codified in the sign of its energy gap: $\Omega>0$ corresponds to excitations, $\Omega<0$ corresponds to de-excitations.

We let the detector interact for an infinite proper time with the Klein-Gordon field, whose initial state is either the ground state or a thermal state at inverse-temperature $\beta$ with $(\gamma,v_\kappa)$-boundary conditions, see Equation \eqref{eq: two point function ground and thermal psi}. In addition, we pa\-ram\-e\-triz\-e the trajectory at fixed spatial coordinates $(r,\theta,\varphi)$ by its proper time $\tau=t/r$. In this setting, letting $s := \tau - \tau'$, the transition rate reads \cite{Fewster2016ewy,Louko2007mu}
\begin{align}
	\label{eq:transition rate}
	\dot{\mathcal{F}}_{\kappa} &= \int_\mathbb{R}ds\, e^{-i\Omega s} \omega_{2,\kappa}\left(r\, \tau,r,\theta,\varphi,r\,\tau',r,\theta,\varphi\right).
\end{align}

Using the two-point functions as per Equation \eqref{eq: two point function ground and thermal time function}, as well as the addition formula between spherical harmonics, Equation \eqref{eq:transition rate} yields
\begin{align*}
	\dot{\mathcal{F}}_{\kappa} &= C_\beta(\Omega) \sum\limits_{\ell=0}^{\infty}(2\ell+1) \frac{ \psi_{\kappa}(r)^2 }{2r\mathcal{N}_\kappa}\bigg|_{p = \frac{|\Omega|}{r}},
\end{align*}
where, being $\Theta$ the Heaviside step function,
\begin{align}
	\label{eq: def Cbeta}
C_\beta(\Omega)  :=
\begin{cases}
\Theta(-\Omega), & \text{for } \beta=\infty,\\
\frac{\sign\Omega}{e^{\beta  \frac{\Omega }{r} } - 1} & \text{for } \beta<\infty.
\end{cases}
\end{align}
\noindent Observe that the variable $ p$ must be evaluated at $\frac{|\Omega|}{r}$, but, for simplicity, we shall not indicate it explicitly in the following.


Recalling that only the $\ell=0$ term calls for a boundary condition, let us decompose the transition rate as
\begin{align*}
\dot{\mathcal{F}}_{\kappa} = \dot{\mathcal{F}}_\kappa^{(\gamma)} + \dot{\mathcal{F}}^{\text{(c)}}.
\end{align*}
The component
\begin{align}
	\label{eq: def trans kappa gamma}
\dot{\mathcal{F}}_\kappa^{(\gamma)} := C_\beta(\Omega)  \frac{\psi_{\kappa}(r)^2}{2r\mathcal{N}_\kappa}
\end{align}
corresponds to the $\ell=0$ contribution, while the remainder
\begin{align}
	\label{eq: trans common term}
\dot{\mathcal{F}}^{\text{(c)}}:=  C_\beta(\Omega)  \sum_{\ell=1}^\infty \frac{2\ell+1}{4r} u ^2
\end{align}
is independent of both $\kappa$ and $\gamma$.

There are two scenarios, i) and ii) as described in the following, where we can take the sum over $\ell$ giving an analytic expression for $\dot{\mathcal{F}}_{\kappa}$.
\begin{itemize}
\item[i)]\label{page item i}
		If $m_0=0$, then $\nu=\ell+1/2$ and thus $\nu=1/2$ for $\ell=0$. Consequently $\dot{\mathcal{F}}_1 \equiv \dot{\mathcal{F}}_2$ and, using the identity
\begin{align*}
  \sum_{\ell=0}^\infty (2\ell+1)j_\ell(z)^2 = 1,
\end{align*}
we obtain
\begin{align}
	\label{eq: m0 gammaNOTnecessarily0 transition rate}
	\dot{\mathcal{F}}^{\text{(c)}} &= C_\beta(\Omega)   r\left[\frac{p}{2\pi}  - \frac{1}{2\pi}  \frac{\sin^2(pr)}{p r^2}\right].
\end{align}

\item[ii)]
		If $\gamma=0$, the secondary solution plays no r\^{o}le and it holds true that $\dot{\mathcal{F}}_1 \equiv \dot{\mathcal{F}}_2$.
\end{itemize}
Note that if either i) or ii) holds true, then $ \dot{\mathcal{F}}_1 \equiv \dot{\mathcal{F}}_2 =: \dot{\mathcal{F}}$. Moreover, if both hold true, then using the expression obtained in item i) above, we get
$
	\dot{\mathcal{F}} = C_\beta(\Omega)  \frac{|\Omega|}{2\pi},
$ which is consistent with the fact that $\text{AdS}_2 \times \mathbb{S}^2$ is conformal to Minkowski spacetime.

\subsection{The $\ell=0$ contribution}

In this section we focus on the $\ell=0$ term: $\dot{\mathcal{F}}_\kappa^{(\gamma)}$. 
First, note that the secondary solutions $v_1$ and $v_2$, as per Equation \eqref{eq: sec sol v1 v2}, are markedly different when $\nu$ is close to $1$, that is for $m_0\sim \sqrt{\frac{3}{4}}$. Let us thus focus in this section on this mass range. We shall show that, for $\gamma\in [0,\frac{\pi}{2}]$ and $\nu_0\sim 1$, the detector behaves in a qualitatively different way for the various choices of secondary solution. We can verify it analytically and visualize it numerically, as shown next. As a matter of fact expanding  $\dot{\mathcal{F}}_\kappa^{(\gamma)}$ as $\nu_0\sim 1$ yields
\begin{widetext}
\begin{subequations}
\label{eq: an exp F1gamma F2gamma nu0 1}
\begin{align}
	\dot{\mathcal{F}}_1^{(\gamma)} = & C_\beta(\Omega) \bigg\{ \nonumber \\
	& (\nu_0 -1)^0\cdot \frac{1}{4}\frac{\left(r \cos (\gamma ) J_1(|\Omega| ) - |\Omega|  \sin (\gamma ) Y_1(|\Omega|)\right)^2}{r^2\cos ^2(\gamma )+|\Omega|^2 \sin ^2(\gamma )} + \nonumber \\
        + &(\nu_0 -1)^1\cdot \frac{1}{2} \frac{ (r \cos (\gamma ) J_1(| \Omega | )-| \Omega |  \sin (\gamma ) Y_1(| \Omega | )) \left(r \cos (\gamma ) \left[\partial_{\nu_0} J_{\nu_0} (| \Omega | )|_{\nu_0=1}\right] - |\Omega |  \sin (\gamma )  \left[\partial_{\nu_0} Y_{\nu_0} (| \Omega | )|_{\nu_0=1}\right]\right)}{ r^2 \cos ^2(\gamma ) + |\Omega|^2  \sin ^2(\gamma ) } + \nonumber \\
				+& o(\nu_0-1)^2\bigg\},\label{eq: an exp F1gamma F2gamma nu0 1a} \\
	\dot{\mathcal{F}}_2^{(\gamma)} = & C_\beta(\Omega) \bigg\{ \nonumber\\& (\nu_0 -1)^0\cdot 	\frac{1}{4}  J_1(|\Omega| )^2 + \nonumber \\
	+& (\nu_0 -1)^1 \cdot  \frac{1}{2} \frac{  J_1(| \Omega | ) \left(r^2 \cos (\gamma \left[\partial_{\nu_0} J_{\nu_0} (| \Omega | )|_{\nu_0=1}\right]-| \Omega | ^2 \sin (\gamma )  \left[\partial_{\nu_0} J_{\nu_0} (| \Omega | )|_{\nu_0=-1}\right] \right)}{ r^2 \cos ^2(\gamma ) - |\Omega|^2  \sin ^2(\gamma ) } + \nonumber\\
	+& o(\nu_0-1)^2\bigg\}.\label{eq: an exp F1gamma F2gamma nu0 1b}
\end{align}
\end{subequations}
\end{widetext}



\begin{widetext}
	\begin{figure}\begin{minipage}{\textwidth}
			\caption{\label{fig: l0 term} The $\ell=0$ contribution to the transition rate at $r=3$ for a massive field with $m_0=\sqrt{\frac{3}{4}}-10^{-2}$ at inverse-temperature $\beta=2\pi$ admitting the generalized $(\gamma,v_1)$- and $(\gamma,v_2)$-boundary conditions, respectively, on the left and on the right columns. On top, seen as a function of $\Omega$, each curve corresponds to a fixed value of $\gamma$. In the middle, density plots for varying both $\gamma$ and $\Omega$. On the bottom, the same density plots of the middle, but with dashed vertical (white) lines corresponding to the first four zeros $j_i$, $i\in\{1,2,3,4\}$, of $J_1(\Omega)$. At the bottom right, the wiggling, dashed, red line is given by $|\Omega| = r \cot^{\frac{1}{2\nu_0}}(\gamma) $. The three points marked by a circle, a triangle and a square on image (b) map to those of image (f). All images share the $\Omega$-axis at the bottom.}
			\scriptsize
			\begin{tabular}{c c c}
				\hspace{-10pt}\includegraphics[align=c,width=.41\textwidth]{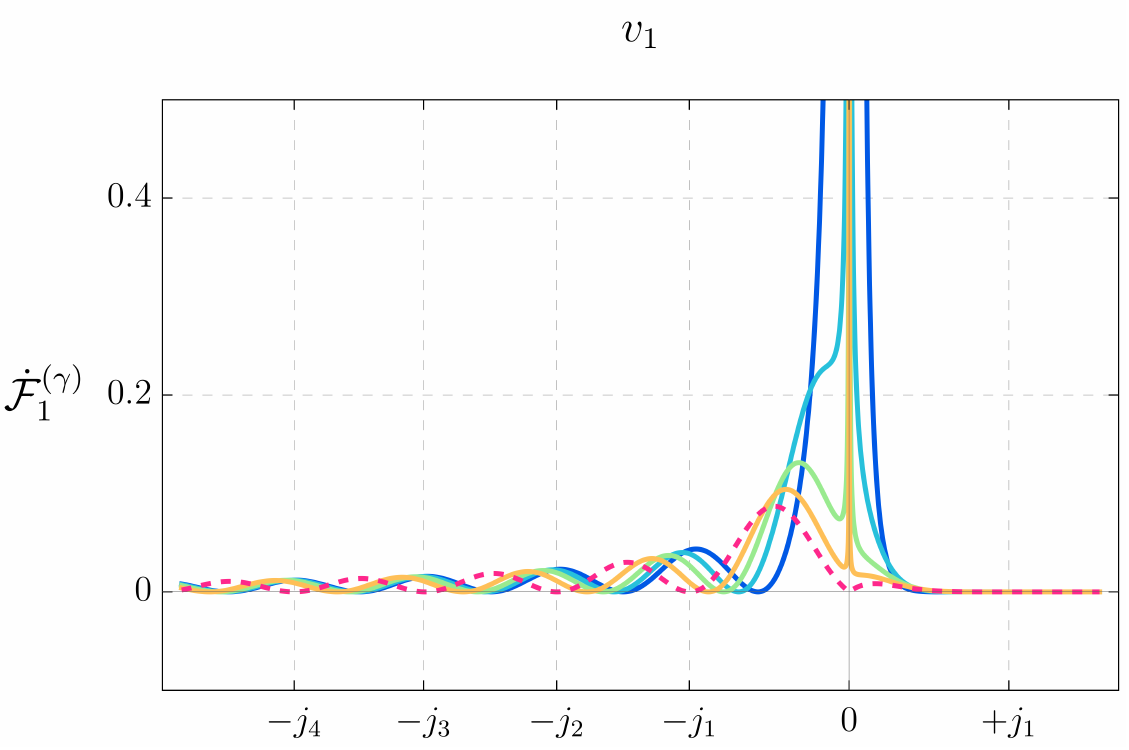}    			 &
				\hspace{-10pt}\includegraphics[align=c,width=.41\textwidth]{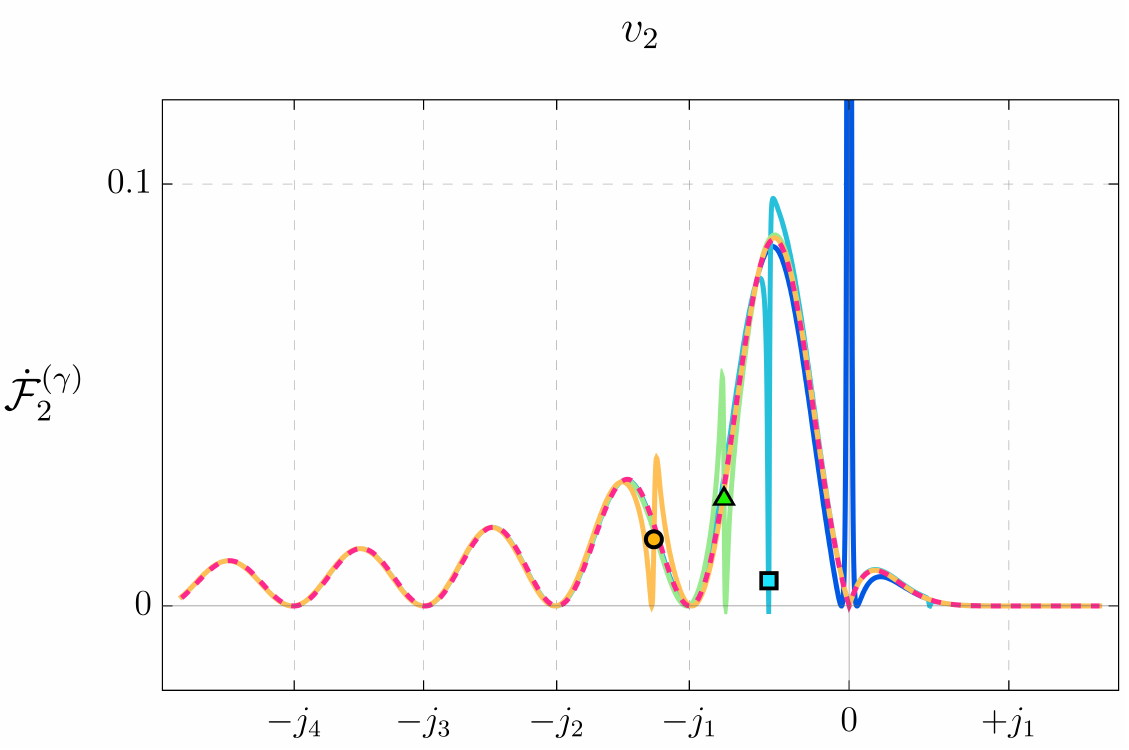}    			 &
				\includegraphics[align=c,width=.08\textwidth]{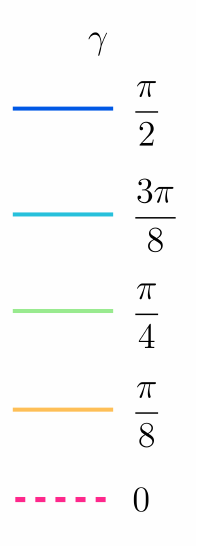}  			 \\
				%
				%
				\hspace{.65cm}(a) & \hspace{.65cm}(b) &  \\

				\includegraphics[align=c,width=.4\textwidth]{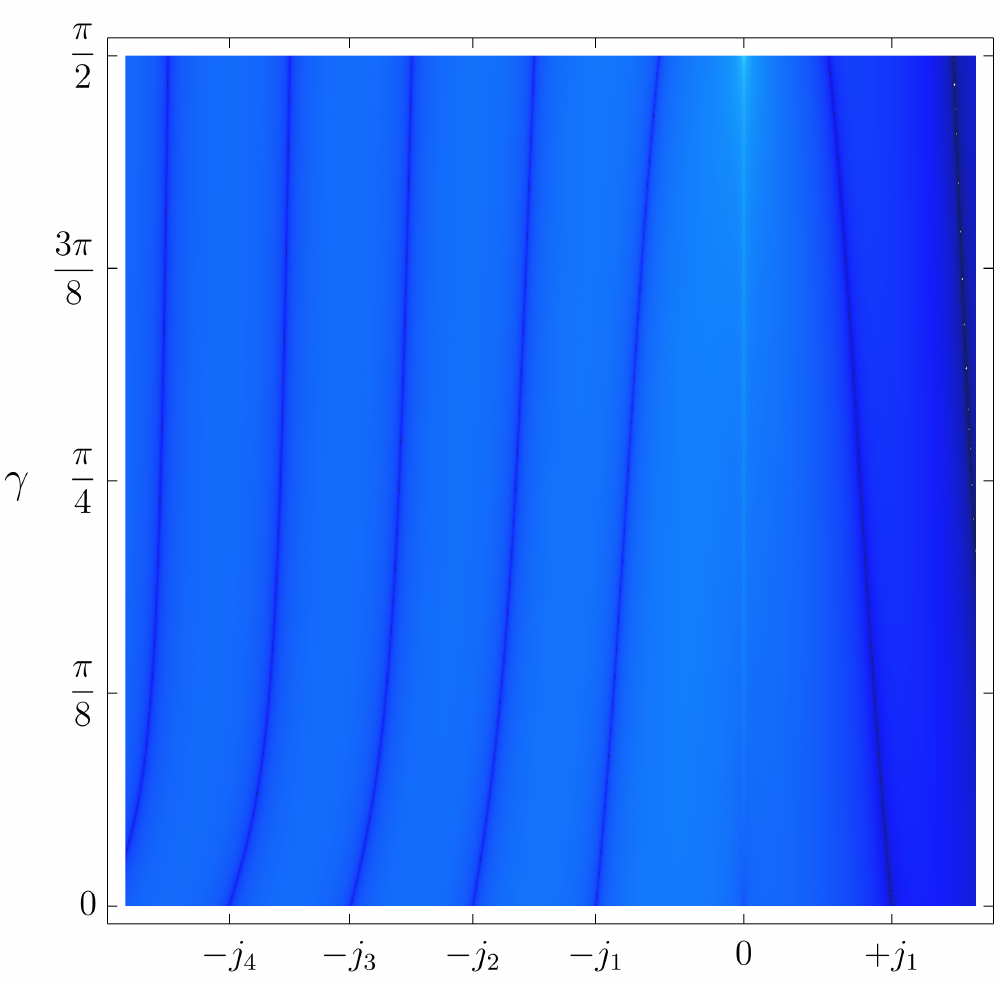}        			   &
				\includegraphics[align=c,width=.4\textwidth]{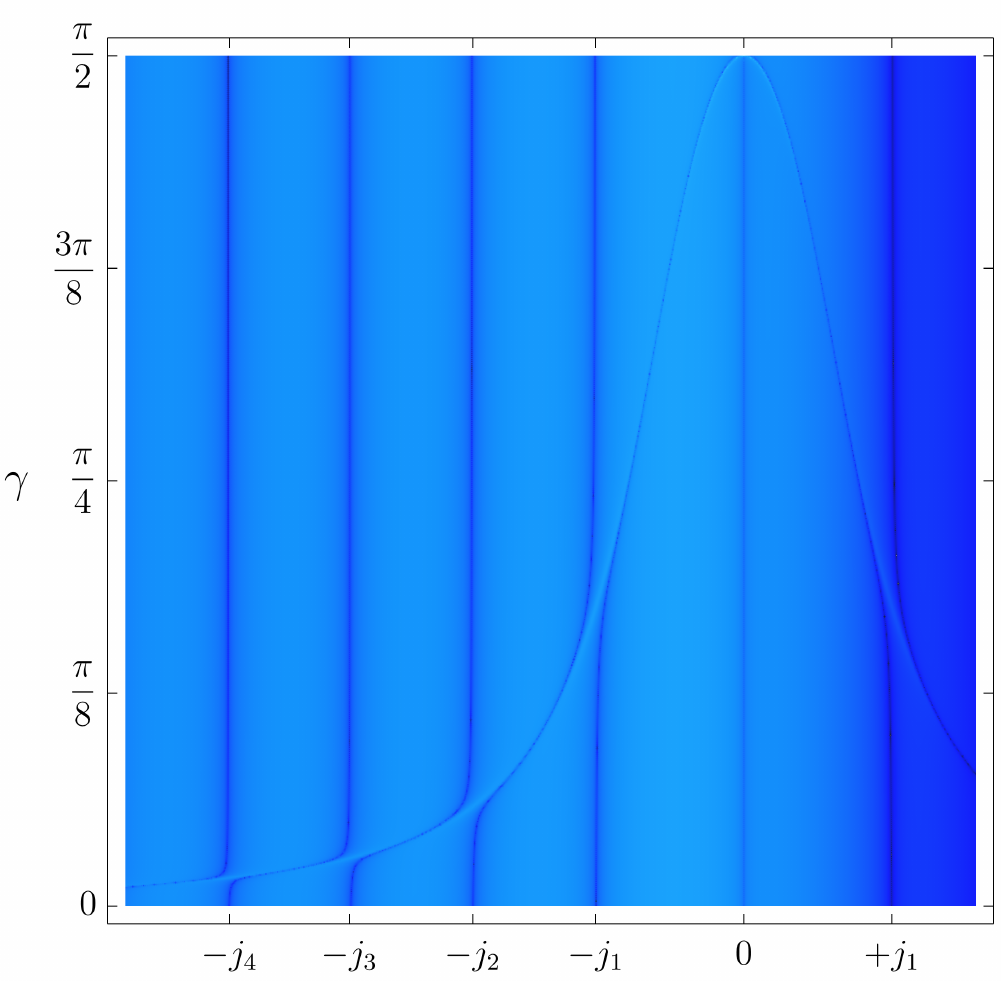}        			   &
				\includegraphics[align=c,width=.08\textwidth]{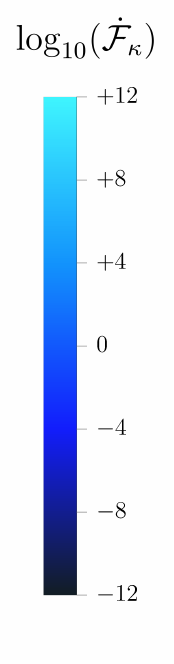}  			 \\
				%
				%
				\hspace{.65cm}(c) & \hspace{.65cm}(d) &  \\

				\includegraphics[align=c,width=.4\textwidth]{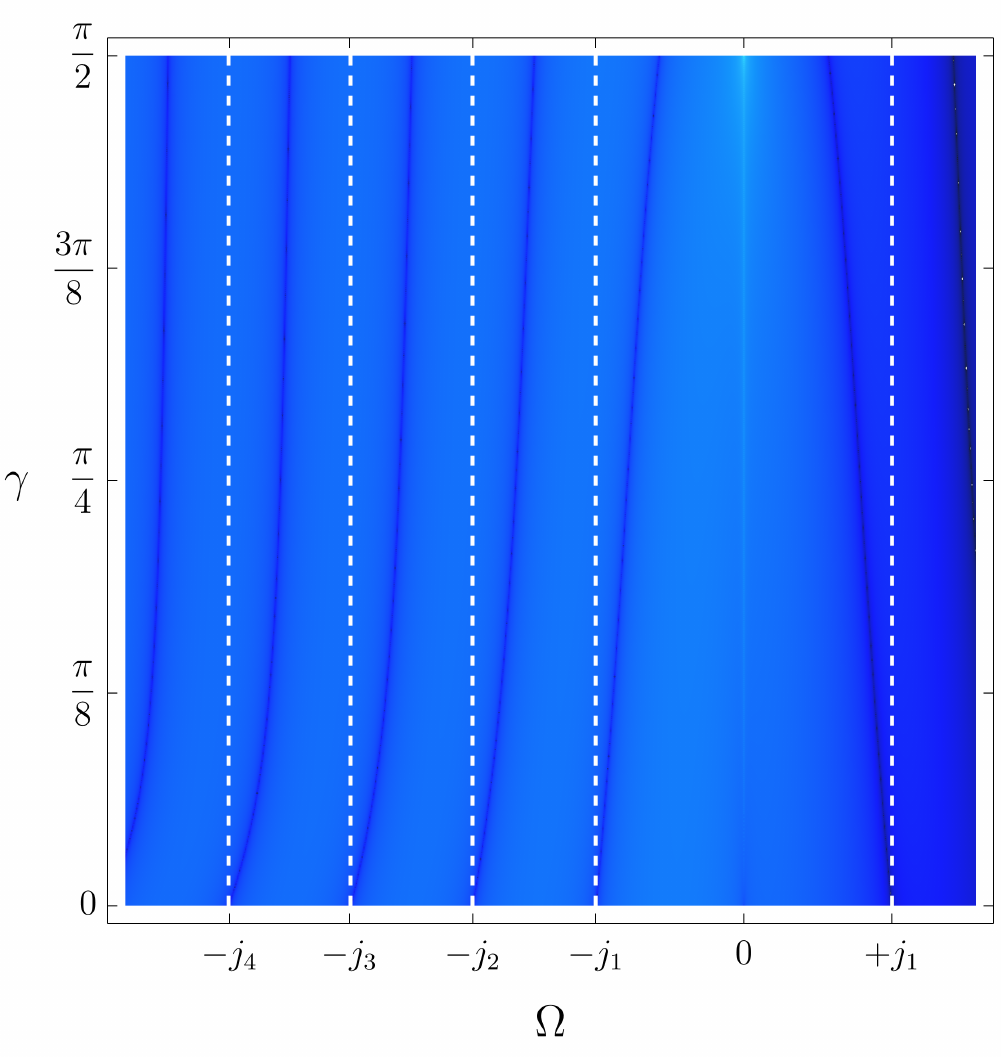} &
				\includegraphics[align=c,width=.4\textwidth]{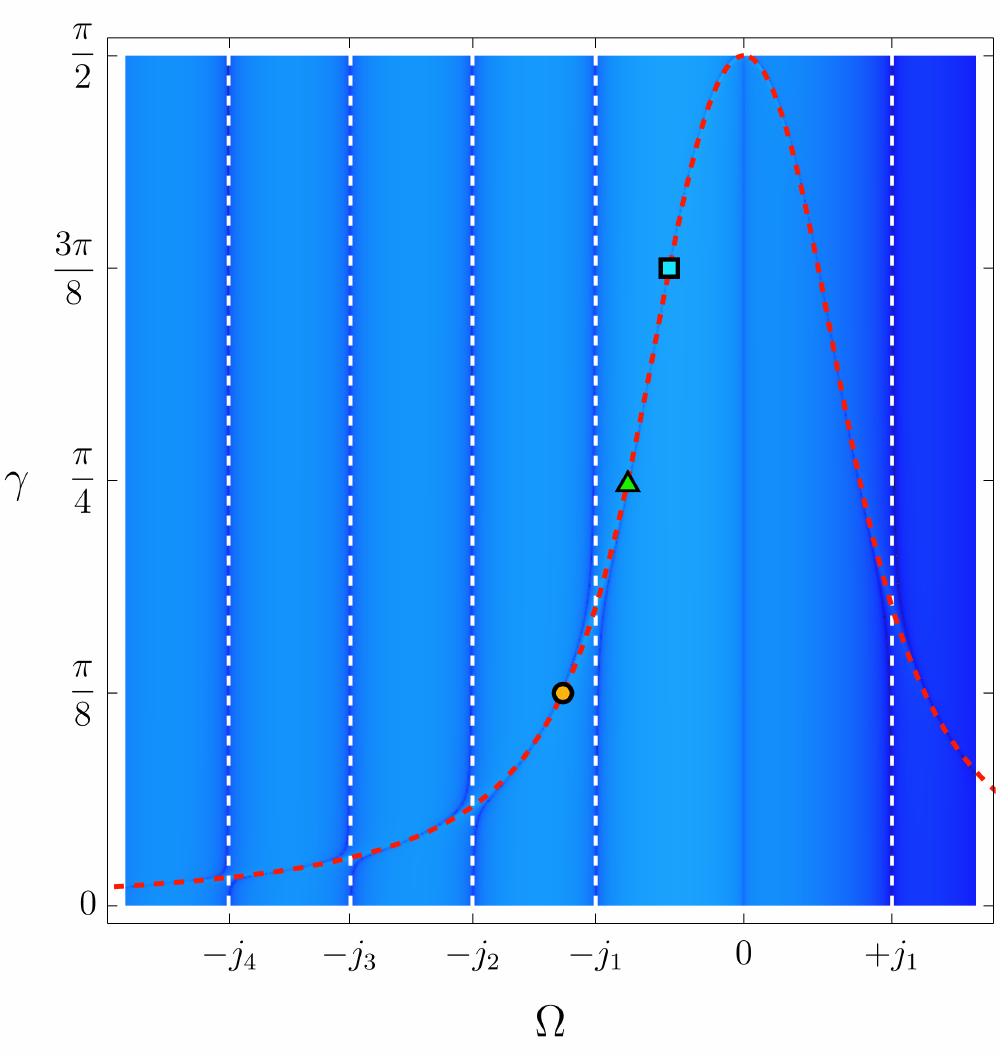} &
				\begin{minipage}{.08\textwidth}
					\setlength{\extrarowheight}{7pt}
					\scriptsize	\vspace{-.4cm}
					\begin{tabular}{c}
						Zeros of $J_1$\\
						$j_1\sim 3.83$ \\
						$j_2\sim 7.01$ \\
						$j_3\sim 10.2$ \\
						$j_4\sim 13.3$ \\ \\ \\
					\end{tabular}
					\includegraphics[width=1.6\textwidth]{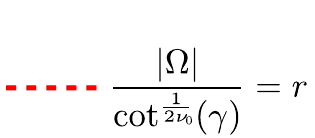}
				\end{minipage}\\
				%
				%
				\hspace{.65cm}(e) & \hspace{.65cm}(f) &  \\
				& & \\
			\end{tabular}
	\end{minipage}\end{figure}
\end{widetext}

\noindent The consequences of Equation \eqref{eq: an exp F1gamma F2gamma nu0 1} are two-fold:
\begin{itemize}
	\item[1)] at zeroth order, the zeros of $\dot{\mathcal{F}}_2^{(\gamma)}$ are those of $J_1(|\Omega|)$ and they do not depend on $\gamma$, while those $\dot{\mathcal{F}}_1^{(\gamma)}$ do;
	\item[2)] by direct inspection of Equation \eqref{eq: an exp F1gamma F2gamma nu0 1b}, $|\Omega| = r \cot^{\frac{1}{2\nu_0}}(\gamma) $ is a pole of $\dot{\mathcal{F}}_2^{(\gamma)}$, while Equation \eqref{eq: an exp F1gamma F2gamma nu0 1a} entails that no pole occurs if $\kappa=1$.
\end{itemize}
These features are manifest in Figure \ref{fig: l0 term} which is displayed at the end of the paper for layout purposes. It contains line and contour plots of  $\dot{\mathcal{F}}_1^{(\gamma)}$ and of $\dot{\mathcal{F}}_2^{(\gamma)}$, respectively in the left and right columns. This numerical evaluation was performed using the software Mathematica and it is presented in a notebook available online \cite{github_lsc}. For this analysis, we considered a field at inverse-temperature $\beta=2\pi$ and of mass $m_0=\sqrt{\frac{3}{4}}-10^{-2}$, which corresponds to $\nu_0 \approx 0.99$. Consistently with the asymptotic expansion in Equation \eqref{eq: an exp F1gamma F2gamma nu0 1}, Figure \ref{fig: l0 term} displays the following notable features:
\begin{itemize}
	\item The dashed, vertical white lines of subfigures (e) and (f) in Figure \ref{fig: l0 term} correspond to the zeros of $J_1$. These lines align with the zeros of $\dot{\mathcal{F}}_2^{(\gamma)}$ for all $\gamma\in\left[0,\frac{\pi}{2}\right]$ as shown in subfigure (f), but this is not the case for $\kappa=1$, as shown in subfigure (e). Accordingly, we refer to this difference in behavior between $\dot{\mathcal{F}}_1^{(\gamma)}$ and $\dot{\mathcal{F}}_2^{(\gamma)}$ as a {\em zeroth-order high-mass effect}.
	\item The dashed, peaked, red line of subfigure (f) of Figure \ref{fig: l0 term} codifies the identity $|\Omega| = r \cot^{\frac{1}{2\nu_0}}(\gamma) $. The restriction of this curve to subfigure (b) is given by the three highlighted points: the circle, the triangle and the square. We can see that these points indicate a {\em first-order high-mass effect} manifest only for $\kappa=2$, and that brings about an extra zero for $\dot{\mathcal{F}}_2^{(\gamma)}$ that depends on the choice of $\gamma$.
\end{itemize}


\normalsize


\subsection{The sum over $\ell$}

\begin{widetext}
\begin{figure}[H]
  \centering
\includegraphics[align=t,width=.12\textwidth]{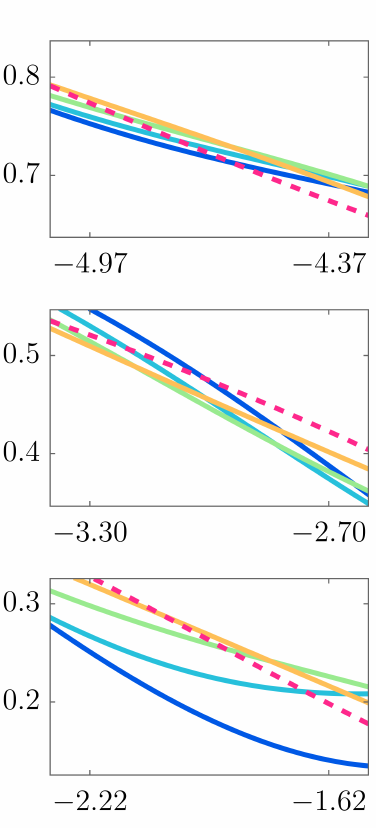}
	 \includegraphics[align=t,width=.35\textwidth]{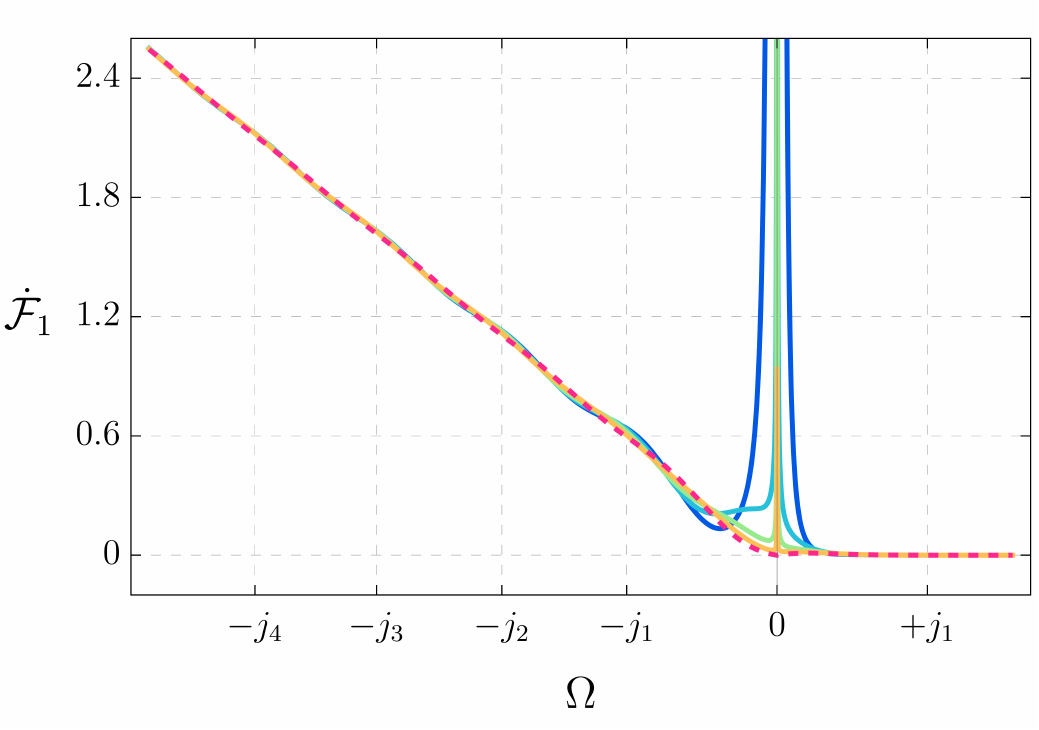}    \includegraphics[align=t,width=.35\textwidth]{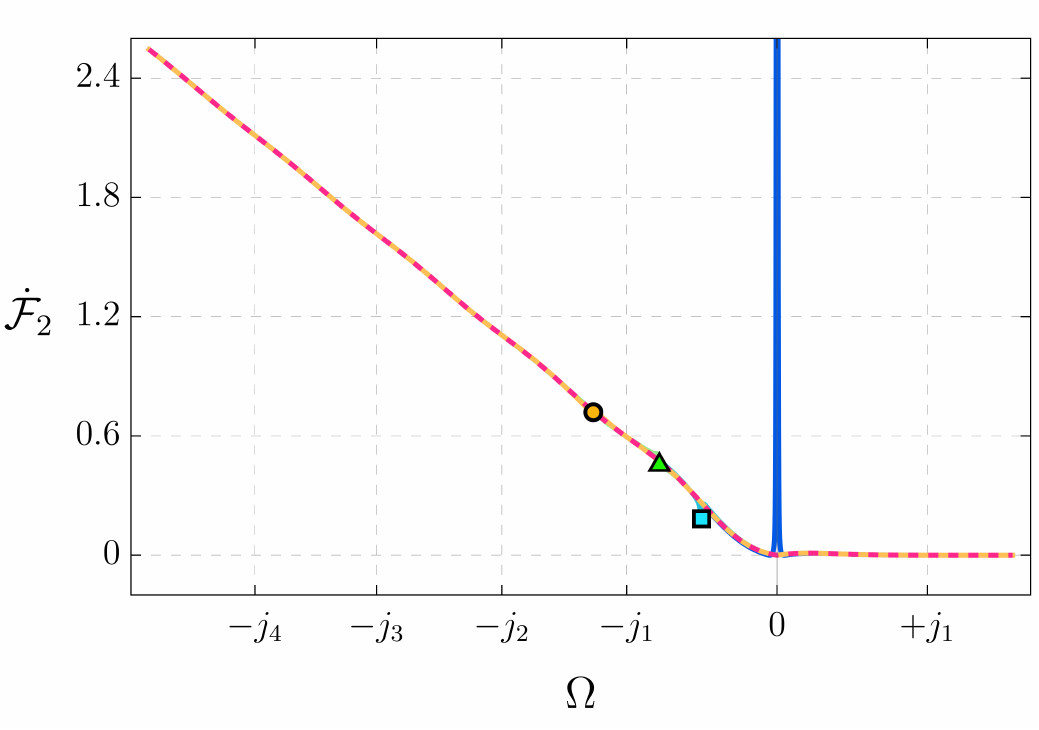} \includegraphics[align=t,width=.12\textwidth]{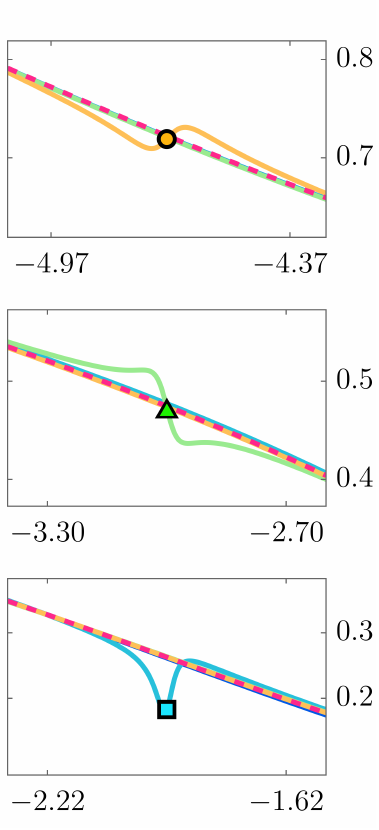}
  \caption{The transition rate summed up to $\ell_{\mini{\text{max}}}=100$, with the same parameters and legends as those of Figure \ref{fig: l0 term}. On the left, for $\kappa=1$. On the right, for $\kappa=2$. The three small plots on the right side provide a zoom-in to a neighborhood of the points highlighted by a circle, a triangle and a square. The three small plots on the left correspond to a zoom-in to the same region of those on the right. The contrast observed between the two sets of three small plots is due to the {\em high-mass effects}.}
  \label{fig:summed up to 100}
    \end{figure}
\end{widetext}

The sum over $\ell$ conceals the high-mass effect discussed in the previous section, see Figure \ref{fig:summed up to 100} that shows the transition rate summed up to $\ell_{\mini{\text{max}}}=100$ with the same parameters of Figure \ref{fig: l0 term}. Nonetheless we can still disambiguate between the choices of secondary solution for the radial equation if, instead of checking out its zeros, we focus on the behaviour of the transition rate at  $\Omega=0$. Figure \ref{fig:summed up to 100} corroborates this statement, but it does not clarify what is happening at such locus. Yet, we can analytically compute the limit of $\dot{\mathcal{F}}_\kappa$ for vanishing energy gap, as we show in the following.  

For $\ell\geq1$, we have that $\nu> \frac{1}{2}$ regardless of the value of the field mass. Thus, using the following two identities
\begin{align*}
	&\lim\limits_{\Omega\rightarrow 0} J_\nu(\Omega) = 0, \text{ for }\nu>0,\\
  & \lim\limits_{\Omega\rightarrow 0} \frac{J_\nu^2(\Omega)}{e^{\Omega}-1} = 0, \text{ for }\nu>\frac{1}{2},
\end{align*}
one can show that
	\begin{equation*}
	 \lim\limits_{\Omega\rightarrow 0} \dot{\mathcal{F}}^{\text{(c)}} = 0, \text{ for }\beta\leq \infty.
	\end{equation*}
Consequently, the behavior of $ \dot{\mathcal{F}}_\kappa$ at $\Omega=0$ depends only on the contribution from the $\ell=0$ term: $\dot{\mathcal{F}}_\kappa^\gamma$, defined in Equation \eqref{eq: def trans kappa gamma}.

For vanishing mass, $m^2_0=0$, $\dot{\mathcal{F}}:=\dot{\mathcal{F}}_1\equiv \dot{\mathcal{F}}_2$. Since
\begin{align*}
  & \lim\limits_{\Omega\rightarrow 0} \frac{J_{\nu}^2(\Omega)}{e^{c\Omega}-1} = \frac{2}{\pi c}, \text{ for }c>0, \, \nu=\frac{1}{2},
\end{align*}
we obtain that, for $\gamma\in\left[0,\frac{\pi}{2}\right)$,
				\begin{equation}
					\label{eq: lim transition m0 gamma0}
			 		\lim\limits_{\Omega\rightarrow 0} \dot{\mathcal{F}} =
					\begin{cases}
						0, &\text{ for }\beta=\infty;\\
						\frac{(r+\tan(\gamma))^2}{2\pi r\beta}, &\text{ for }\beta<\infty.
					\end{cases}
				\end{equation}

Observe that if $\gamma=\pi/2$ then $\lim\limits_{\Omega\to 0}\dot{\mathcal{F}}$ diverges for $\beta\leq\infty$.
For positive masses, $\dot{\mathcal{F}}_1 \not\equiv \dot{\mathcal{F}}_2$ and the results are summarized in Table \ref{tab: limits as Omega 0 for m greater than 0}.

\renewcommand{\arraystretch}{1.5}
\begin{table}[H]
   \centering
	\caption{The limits for $m_0\in\left(0,\sqrt{\frac{3}{4}}\right)$.}
   \begin{tabular}{c c c c}
   \toprule
	 State &
   $\gamma$&
   $\lim\limits_{\Omega\rightarrow 0}\dot{\mathcal{F}}_1 $&
   $\lim\limits_{\Omega\rightarrow 0}\dot{\mathcal{F}}_2$ \\
	 \hline\hline
      \multirow{ 2}{*}{$ \beta=\infty$ } &  $\left[0,\frac{\pi}{2}\right)$ & $0$      &  $0$       \\
			                          &  $\frac{\pi}{2}$                & $\infty$ &  $\infty$  \\\hline
	  \multirow{ 3}{*}{$ \beta<\infty$ } &  $0$                            & $0$      &  $0$       \\
		\rowcolor{cyan} $ \beta<\infty$     &  $\left(0,\frac{\pi}{2}\right)$ & $\infty$ &  $0$       \\
		                           &  $\frac{\pi}{2}$                & $\infty$ &  $\infty$  \\
 \bottomrule
   \end{tabular}
\label{tab: limits as Omega 0 for m greater than 0}
\end{table}



All in all, the boundary condition chosen for the mode $\ell=0 $ affects the behavior of the transition rate in its full form, both quantitatively and qualitatively. In fact, its limit at $\Omega\rightarrow 0$ receives no contribution from the terms with $\ell>0$, as shown above. Markedly, the transition rate of an Unruh-DeWitt detector interacting with a Klein-Gordon field at a finite temperature state, with $m_0\in\left(0,\sqrt{\frac{3}{4}}\right)$, and admitting a generalized $(\gamma,v)$-boundary condition with $\gamma\in\left(0,\frac{\pi}{2}\right)$ explicitly discriminates between $\kappa=1$ and $\kappa=2$ in Equation \eqref{eq: sec sol v1 v2}, as highlighted in blue in Table \ref{tab: limits as Omega 0 for m greater than 0}.


\section{Outlook}
\label{sec: Summary}


We have shown that choosing one among the generalized $(\gamma,v)$-boundary conditions is not a mere mathematical exercise, rather it has notable physical consequences. For definiteness we have considered a Klein-Gordon field on $\text{AdS}_2 \times \mathbb{S}^2$ spacetime and we have constructed the two-point correlation function of the ground and thermal states for two families of boundary conditions, characterized by different choices of secondary solution, $v_1$ and $v_2$, see Equation \eqref{eq: sec sol v1 v2}. We probed the underlying system with an Unruh-DeWitt detector and we showed that its transition rate differentiates between the two families of boundary conditions.

There are two notable facts that highlight the hidden freedom in the mode expansion of the Klein-Gordon field and connect our results to \cite{Campos2022byi}. First, the boundary conditions with $(\gamma,v_1)$ and $(\gamma,v_2)$ are equivalent at the level of the radial differential equation, but cease to be such at the level of the whole Klein-Gordon equation. This is hinted by the map that relates the parameter $\gamma$ in the two settings. It depends on the Fourier frequency associated to the time coordinate and, therefore, we can map a solution satisfying Equation \eqref{eq: robin bc} for $\kappa =1$, into one satisfying it for $\kappa=2$ via
	\begin{equation*}
		\tan(\gamma) \mapsto  \frac{p^{2\nu}\sin(\pi\nu)}{p}\frac{\tan(\gamma)}{1+\tan(\gamma)p^{2\nu}\cos(\pi\nu)}.
	\end{equation*}
There is an explicit dependence of the Fourier parameter $p$ which in turn related to the underlying frequency. This entails that, while, at the level of radial equation, this is a mere algebraic relation, from a fully covariant viewpoint, this is no longer true as one can realize by an inverse Fourier transform. Finally, our work sets the ground for multiple future investigations that are worth mentioning:
\begin{itemize}
	\item comparisons between the physical phenomena associated to different choices of secondary solution, possibly considering other observables such as the stress-energy tensor;
	\item proving the universality of the features highlighted in this paper by studying their occurrence in other globally hyperbolic spacetimes with a timelike boundary, such as for example global monopoles and cosmic strings;
	\item studying the influence of the $(\gamma,v)$-boundary conditions on specific phenomena such as the anti-Hawking effect, see \cite{DeSouzaCampos2020ddx, Brenna:2015fga, Robbins:2021ion, deSouzaCampos:2020bnj};
	\item a further investigation of the behaviour of the transition rate at $\Omega\rightarrow 0$ particularly in connection with the notion of a structureless scalar source \cite{Cozzella2020gci}.
\end{itemize}


\section{Acknowledgments}
The work of L.C. is supported by a postdoctoral fellowship of the Department of Physics of the University of Pavia, while that of L.S. by a PhD fellowship of the University of Pavia.


\end{multicols}


\end{document}